\documentclass[conference, a4paper]{IEEEtran}
\IEEEoverridecommandlockouts
\usepackage{cite}

\ifCLASSINFOpdf
  \usepackage[pdftex]{graphicx}
  \graphicspath{{../pdf/}{../jpeg/}}
   \DeclareGraphicsExtensions{.pdf,.jpeg,.png}
\else
\fi
\usepackage[cmex10]{amsmath}
\usepackage{xfrac}
\usepackage{array}

\usepackage{mdwmath}
\usepackage{amsfonts}
\usepackage{amssymb}
\usepackage{amsbsy}
\usepackage{amsthm}

\begin{document}

%
\title{Coverage Region Analysis for MIMO Amplify-and-Forward Relay Channel with the\\ Source to Destination Link}



%
\author{
\IEEEauthorblockN{Behrooz Razeghi,
Ghosheh Abed Hodtani,
Seyed Alireza Seyedin}
Department of Electrical Engineering, Faculty of Engineering\\
Ferdowsi University of Mashhad,
Mashhad, Iran\\
Email: behrooz.razeghi.r@ieee.org, ghodtani@gmail.com, seyedin@um.ac.ir
}

\maketitle

\begin{abstract}
In this paper we study and analyze coverage region for half-duplex multiple-input multiple-output (MIMO) relay channel with amplify-and-forward (AF) strategy at the relay station. By assuming mixed Rayleigh and Rician fading channels with two different relay station situations, we consider the objective of maximizing coverage region for a given transmission rate and find the optimal relay location in the sense of maximizing coverage region. Using Monte Carlo simulations, the coverage region and capacity bounds are shown for different fading cases and different relay station locations. Finally, we compare our results with previous ones obtained for decode-and-forward (DF) MIMO relay channel.

\end{abstract}


\begin{IEEEkeywords} Optimal relay location; coverage region; MIMO relay channel; amplify-and-forward strategy; desired transmission rate.
\end{IEEEkeywords}

%
\IEEEpeerreviewmaketitle

\vspace{2pt}
\section{Introduction}

In wireless cooperation networks, the relay channel is probably the most fundamental structural unit and relay technologies have attracted substantial research interest since they can improve communications capacity and performance, expand coverage region, and reduce battery consumption. Among relaying strategies, amplify-and-forward (AF) strategy is the most commonly used relaying strategies because of its simplicity in implementation.

Van der Meulen \cite{van1971three} first introduced the relay channel. In \cite{cover1979capacity}, Cover \& El-Gamal studied it in detail, where they proved capacity of degraded and reversely degraded, Gaussian degraded and full feedback relay channel as well as general capacity upper and lower bounds. 
In \cite{hodtani2008capacity} and \cite{hodtani2009unified} previously introduced capacity theorems for the relay channel have been unified into one capacity theorem.

In \cite{kramer2005cooperative}, the authors studied cooperative strategies in Gaussian relay networks and considered the impact of relocating the relay node on achievable rates at the destination node. However, in actual situations the location of the relay node is determined by network designers at the time of network design and the objective problem is to maximize coverage area for a given transmission rate. In \cite{aggarwal2009maximizing} the authors studied the problem of maximizing coverage region for a given rate and evaluated decode-and-forward (DF) and compress-and-forward (CF) strategies with the objective of maximizing coverage region for Gaussian point to point relay channel. 
%

In \cite{alizadeh2012analysis} and \cite{alizadeh2013analysis} the authors analyzed the coverage region in MIMO relay channel with single relay in Rayleigh fading case and determined the optimal relay location maximizing the coverage region. The coverage region for Rayleigh and Rician fading MIMO channels with multiple decode-and-forward relays was studied in \cite{razeghi2014iswcs}, also the authors derived a simple analytical formula for coverage region extension factor in cooperation scenario. In \cite{razeghi2014twohop}, the coverage region in MIMO relay networks with multiple amplify-and-forward relays has been analyzed, also the impact of spatial correlation between transmit and receive antennas was examined.

\emph{Our work}: 
In this paper, we consider a half-duplex amplify-and-forward (AF) MIMO relay channel and investigate the effect of channel fading on the coverage region. Since our goal is to obtain maximum coverage region, we put relays at maximum distance for a certain transmission rate at the source such that if we increase the distance, the relay cannot operate satisfactorily. We assume that the relay amplification factor is fixed.
Also, we consider two  situations for the relay station: (1) the situation in which the relay is located above the rooftop (source-relay line-of-sight); (2)  the situation in which the relay is located below the rooftop (relay-destination line-of-sight). Finally, we provide numerical results and compare them with those obtained in \cite{alizadeh2012analysis}.

\begin{figure}[!t]
    \centering
    \includegraphics[scale=1.02]{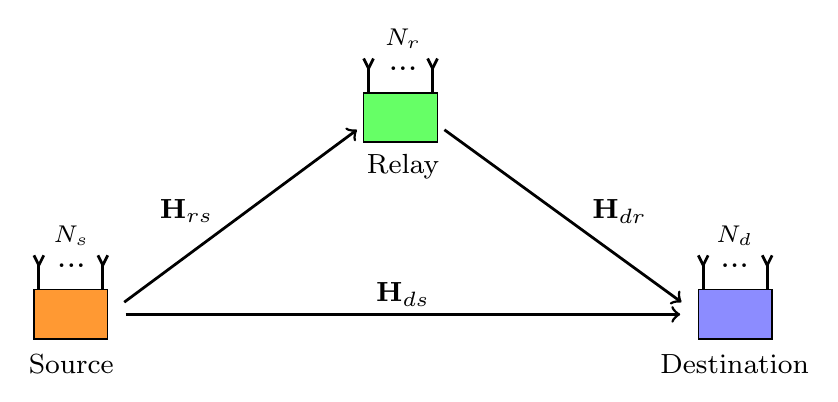}
    \caption{MIMO relay channel.}
    \label{Fig3}
\end{figure}
\vspace{4pt}

\emph{Notations}: Throughout this paper, we use $\mathbb{E}\left\{.\right\}$ to denote the expectation operator; the distribution of a circularly symmetric complex Gaussian vector with mean $\mathbf{m}$ and covariance matrix $\mathbf{Q}$ is denoted as $\mathcal{CN}(\mathbf{m},\mathbf{Q})$; "$\dag$" stands for the conjugate transpose; and vectors and matrices are denoted by boldface lower case ($\textbf{x}$) and upper case letters ($\mathbf{X}$), respectively.

The rest of the paper is organized as follows. In Section II, we present our channel model. We review in Section III the capacity theorems of MIMO relay channel. In Section IV, we define the concept of coverage region and evaluate the desired transmission rate in the sense of optimal relay location. Simulation results are presented in Section V, and Section VI contains our conclusion.




\section{Channel Model and Preliminaries}


Consider one source, one relay, and one destination. Our MIMO relay channel is depicted in Fig.~\ref{Fig1}. In our model, the relay node is assumed to be half-duplex. During the first hop, source transmits to the relay and destination and in the second hop the relay forward scaled version of its received signals from the first hop to its destination. Our channel geometry is depicted in Fig.~\ref{Fig2}.  


\subsection{Amplify-and-Forward Channel Model}
The received signals at the destination node in AF scenario after two hops can be written as 
\begin{align}\label{af.model}
    \mathbf{y}_d = \! 
    \left[ \!
       \begin{array}{c}
         \sqrt{P_{ds}}\mathbf{H}_{ds}\\
          \sqrt{P_{dr}} \sqrt{P_{rs}}\, g\, \mathbf{H}_{dr} \mathbf{H}_{rs}
       \end{array} 
       \! \right] \mathbf{x}_s + \!
       \left[\! \! 
        \begin{array}{c}
          \mathbf{z}_{d_1} \\
          \sqrt{P_{dr}}\, g\, \mathbf{H}_{dr} \mathbf{z}_{r} + \mathbf{z}_{d_2}
        \end{array} 
        \! \! \right]\! ,
\end{align}
where
\begin{itemize}
  \item $\textbf{x}_s$ is the $N_s \times 1$ transmitted symbol vector from the source. The source transmitted symbols are chosen to be i.i.d with $\mathbb{E}\left\{\mathbf{x}_s \mathbf{x}_s^{\dag}\right\} = \frac{P_s}{N_s} \mathbf{I}_{N_s}$,
  \item $\mathbf{H}_{rs} \in \mathcal{C}^{N_r \times N_s},\,\mathbf{H}_{ds} \in \mathcal{C}^{N_d \times N_s}$,\, and $\mathbf{H}_{dr} \in \mathcal{C}^{N_d \times N_r}$,\, are normalized  channel gain matrices for the source to relay, source to destination, and relay to destination, respectively, where their entries are assumed to be zero mean circularly symmetric complex Gaussian (ZMCSCG) random variables,
  \item $P_{rs}$, $P_{ds}$, and $P_{dr}$ are average power of source-relay link, source-destination link, and relay-destination link, respectively,
  \item $g$ is amplification factor,
  \item $\textbf{z}_r,\,\textbf{z}_{d_1}$ and $\mathbf{z}_{d_2}$ are independent $N_r \times 1$, $N_d \times 1$ and $N_d \times 1\,$ ZMCSCG noise vectors at relay, destination in first hop, and destination in second hop, respectively, with $\mathbb{E}\left\{\mathbf{z}_r \mathbf{z}_r^{\dag}\right\} \! \! =\!  \! \mathbf{I}_{N_r}$, $\mathbb{E}\left\{\mathbf{z}_{d_1} \mathbf{z}_{d_1}^{\dag}\right\} \! \!  =\!  \!  \mathbf{I}_{N_d}$ and $\mathbb{E}\left\{\mathbf{z}_{d_2} \mathbf{z}_{d_2}^{\dag}\right\}\! \!  =\!   \mathbf{I}_{N_d}$.
\end{itemize}

By defining $F_{{ds}_1}= \sqrt{P_{ds}}$, $F_{{ds}_2}= g \sqrt{P_{dr}} \sqrt{P_{rs}}$, and $F_{{dr}}= g \sqrt{P_{dr}}$, the received signal at the destination node can also be expressed as
\vspace{-2pt}
\begin{equation} \label{combinemodelchannel}
\mathbf{y}_d = \mathbf{H} \mathbf{x}_s + \mathbf{B} \mathbf{n},
\end{equation}
where
\vspace{-2pt}
\begin{equation}
\mathbf{H} = \left[
\begin{array}{c}
F_{{ds}_1} \mathbf{H}_{ds} \\
F_{{ds}_2} \mathbf{H}_{dr} \mathbf{H}_{rs}
\end{array}
\right] , 
\end{equation}
and
\vspace{-2pt}
\begin{equation}
\mathbf{B} = {\left[
\begin{array}{cc}
\mathbf{I}_{N_d} & \mathbf{0} \\
\mathbf{0} & F_{dr}^2 \mathbf{H}_{dr} \mathbf{H}_{dr}^{\dag} +\mathbf{I}_{N_d}
\end{array}
\right]}^{\sfrac{1}{2}} ,
\end{equation}
and $\mathbf{n}$ is normalized noise vector with $\mathbf{n} \sim (\mathbf{0}, \mathbf{I}_{2N_d})$.

The covariance matrix of the input signals can be written as
\begin{equation}
\label{joint.cov}
\mathbf{Q}\triangleq \mathbb{E} \left\{{\left[\mathbf{x}_s \; \mathbf{x}_r\right]}^t { \left[ \mathbf{x}_s \; \mathbf{x}_r\right] }^\dag \right\} =
\begin{bmatrix}
\mathbf{Q_{ss}}& \mathbf{Q_{sr}}\\
\mathbf{Q_{rs}}& \mathbf{Q_{rr}}
\end{bmatrix}.
\end{equation}
Note that $\mathbf{Q}$ is a Hermitian matrix ~\cite{telatar1999capacity}.

%
%
%
\begin{figure}[!t]
    \centering
    \includegraphics[scale=1]{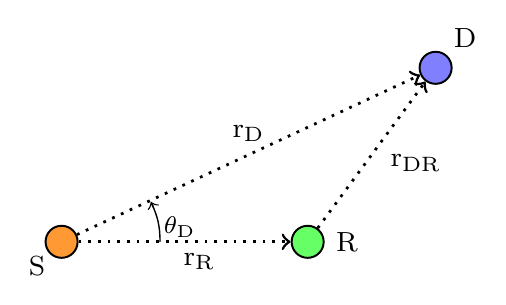}
    \caption{Network geometry.}
    \label{Fig3}
\end{figure}
\vspace{4pt}

\subsection{Line-of-Sight Channel Matrix Model}
We consider two fading channel cases, Rayleigh fading case and Rician fading case. The channel matrix in Rician fading case can be modeled as \cite{paulraj2003introduction,jiang2007modelling,oestges2004propagation}
\begin{align}\label{sr.H}
    \mathbf{H} = \sqrt{\frac{K}{K+1}}\mathbf{{H}}^{\textrm{LOS}} + \sqrt{\frac{1}{K+1}}\mathbf{{H}}^{\textrm{NLOS}},
\end{align}
where
$\sqrt{\sfrac{K}{K+1}}\mathbf{H}^{\textrm{LOS}}$ is line-of-sight component of the channel. $\sqrt{\sfrac{1}{K+1}}\mathbf{H}^{\textrm{NLOS}}$ is non line-of-sight component assuming uncorrelated fading and with consideration of the scattering components and their effect during the propagation. $K$ is the Rician factor for transmitter-receiver channel and is defined as the ratio of the total power in the fixed  (line-of-sight) component of the channel to the power in fading (non line-of-sight) component. $K$ can be modeled as a function of distance between the transmitter and receiver \cite{3GPP2005}. 

The channel matrix for LOS (fixed) component can be modeled as \cite{jiang2007modelling}
\begin{align}\label{los.h}
    \mathbf{H}^{\textrm{LOS}} &=
    \begin{bmatrix}
        \mathbf{\acute H}_{N_iN_j} \odot \mathbf{A}_{N_iN_j}^{VV} & \mathbf{\acute H}_{N_iN_j} \odot \mathbf{A}_{N_iN_j}^{VH}\\
        \mathbf{\acute H}_{N_iN_j} \odot \mathbf{A}_{N_iN_j}^{HV} & \mathbf{\acute H}_{N_iN_j} \odot \mathbf{A}_{N_iN_j}^{HH}
    \end{bmatrix},
\end{align}
where $\odot$ denotes for the element wise multiplication; $\mathbf{\acute H}_{N_iN_j},\, i \in \left\{r, d\right\}, \, j \in \left\{s, r\right\}$ is LOS channel matrix with co-polarized antennas. $\mathbf{A}_{N_iN_j}$ is the matrix representing the polarization mismatch effect in the model. The assumption here is the strict alignment of all the transmit and receive antennas at the source node and relay node; in other words the polarization antennas are not taken into account. Consequently, $\mathbf{A}_{N_iN_j}^{VV}$ and $\mathbf{A}_{N_iN_j}^{HH}$ will be all-one matrices, while $\mathbf{A}_{N_iN_j}^{VH}$ and $\mathbf{A}_{N_iN_j}^{HV}$ will be all-zero matrices. We consider two alternative prototypes of $\mathbf{\acute H}_{N_iN_j}$ which correspond to unwell-conditioned channel and well-conditioned channel in LOS scenario.


\section{Capacity of MIMO Relay Channel}
In this section we review the main capacity results obtained for MIMO relay channel.

\subsection{Cut-set Upper Bound}

The capacity upper bound of a general Gaussian relay channel channel is \cite{cover1979capacity}
\begin{align}\label{cupper}
C_{\mathrm{upper}}\! =\! \! \! \mathop{\max}_{p(x_s,x_r)}\! \! \min \left\{I\! \left(X_s;Y_r,Y_d \mid \! X_r\right)\! , I\! \left(X_s,X_r;Y_d\right) \right\}\! ,
\end{align}
where $I\! \left(X_s;Y_r,Y_d \mid X_r\right)$ corresponds to broadcast (BC) bound, $I\! \left(X_s,X_r;Y_d\right)$ corresponds to multiple-access (MAC) bound, and the maximization is with respect to the joint distribution of the source and relay symbols. Considering $\textbf{x}_i\sim\mathcal{CN}(\mathbf{0},\mathbf{Q}_{ii}), i \in \left\{s,r\right\}$, where $\mathbf{Q}_{ii}$ is the covariance matrix of $\textbf{x}_i$, the mutual information expressions in~\eqref{cupper} can be written for the MIMO relay channel as~\cite{foschini2011opt}  
\begin{align}
\label{Ccs}
C_{\mathrm{CS}} = \max_{\mathbf{Q}_{ii}:\mathrm{tr}(\mathbf{Q}_{ii})\leq P_i,~i=s,r} \min{(C_1,C_2)},
\end{align}
\begin{align}
\label{c1}
C_1 &= \log \det(\mathbf{I}_N+\mathbf{H}_{\mathrm{BC}}\mathbf{Q}_{s \mid r}\mathbf{H}_{\mathrm{BC}}^\dag),
\end{align}
\begin{align}
\label{c2}
C_2 &= \log \det(\mathbf{I}_{N_d}+\mathbf{H}_{\mathrm{MAC}}\mathbf{Q}\mathbf{H}_{\mathrm{MAC}}^\dag),
\end{align}
where $\mathbf{H}_{\mathrm{BC}}=\begin{bmatrix}\mathbf{H}_{ds}\\\mathbf{H}_{rs}\end{bmatrix}$, $\mathbf{H}_{\mathrm{MAC}} = \begin{bmatrix}\mathbf{H}_{ds}&\mathbf{H}_{dr}\end{bmatrix}$, $N=N_r+N_d$
and $\mathbf{Q}_{s|r} \triangleq \mathbb{E}\left\{\mathbf{x}_s\mathbf{x}_s^{\dag}|\mathbf{x}_r\right\}=
\mathbf{Q}_{ss}-\mathbf{Q}_{sr}\mathbf{Q}_{rr}^{-1}\mathbf{Q}_{rs}$ is the conditional covariance matrix and given by Schur complement of $\mathbf{Q}_{rr}$ in $\mathbf{Q}$ ~\cite{boyd2004convex}.
The optimal distribution $p(x_s,x_r)$ in~\eqref{cupper} for Gaussian relay channel is Gaussian~\cite{cover1979capacity}, and consequently the maximization of~\eqref{Ccs} would be with respect to three covariance matrices $\mathbf{Q}_{ss}$, $\mathbf{Q}_{rr}$, and $\mathbf{Q}_{sr}$.

When the entries of channel gain matrices are random and the channel state information is only known at the receivers, the optimal joint transmit covariance matrix $\mathbf{Q}$ in~\eqref{Ccs} is diagonal. The authors in \cite{infotheoricrelay} showed that the optimal solution is
\begin{equation}
\label{cov.matrix}
\mathbf{Q}_{ss}=\frac{P_s}{N_s}\mathbf{I}_{N_s},~~\mathbf{Q}_{rr}=\frac{P_r}{N_r}\mathbf{I}_{N_r},~~
\mathbf{Q}_{sr}=\mathbf{0}.
\end{equation}
This means that the equal power allocation is the optimal solution, where $\mathbf{Q}_{sr}=\mathbf{0}$ refers to the independence between the source and the relay symbols. Therefore, the cut-set upper bound for the half-duplex MIMO relay channel with only CSIR can be expressed as
\begin{equation}
C_{\mathrm{CS}}=\min(C_{1},C_2),
\end{equation}
\begin{equation}
\label{C1}
C_1= \frac{1}{2} \mathbb{E} \left\{\log\det(\mathbf{I}_{N}+\frac{P_s}{N_s}\mathbf{H}_\mathrm{BC} \mathbf{H}_{\mathrm{BC}}^\dag)\right\},
\end{equation}
\begin{equation}
\label{C2}
C_2=\!  \frac{1}{2} \mathbb{E} \left\{ \! \log\det(\mathbf{I}_{N_d}+\mathbf{H}_{\mathrm{MAC}}\! \left[
\begin{matrix}
\frac{P_s}{N_s}.\mathbf{I}_{N_s} &  \mathbf{0}\\
\mathbf{0}                                            &  \frac{P_r}{N_r}.\mathbf{I}_{N_r}
\end{matrix}
\right]\!  \mathbf{H}_{\textmd{MAC}}^\dag)\!  \right\}\! .
\end{equation}
The factor $\sfrac{1}{2}$ refer to transmission in two uses of the channel (half-duplex relay).

\subsection{Amplify-and-Forward Ergodic Capacity}

The ergodic capacity of the amplify-and-forward MIMO relay channel with the source to destination link (direct link) is studied in \cite{firag2009afcapacitydirect}. We recapitulate the main results in this section. The ergodic capacity of the channel model given by (\ref{combinemodelchannel}) can be written as 
\begin{equation}\label{ergodicAFbyH}
C = \frac{1}{2} \mathbb{E} \left\{ \log_2 \det \left( \mathbf{I}_{2N_d} + \frac{P_s}{N_s} \mathbf{H}\mathbf{H}^\dag {\left(\mathbf{B} \mathbf{B}^\dag\right)}^{-1} \right)\right\}.
\end{equation} 
By utilizing the singular value decomposition we can write $\mathbf{H}_{dr} = \mathbf{U}_{dr} \mathbf{D}_{dr} \mathbf{V}_{dr}^{\dag}$, where $\mathbf{D} = \mathrm{diag} \left\{\lambda_1, ..., \lambda_r \right\}$
is $N_d \times N_r$ diagonal matrix with $\left\{\lambda_i\right\}, i=1,..,r$ as the diagonal elements in decreasing order, and $r = \min \left(N_d, N_r\right)$. Also $\mathbf{U}_{dr} \in \mathcal{C}^{N_d \times N_d}$ and $\mathbf{V}_{dr} \in \mathcal{C}^{N_r \times N_r }$ are unitary matrices. Using the identity $\det \left(\mathbf{I} + \mathbf{A}\mathbf{B}\right) = \det \left(\mathbf{I}+\mathbf{B}\mathbf{A}\right)$, the ergodic capacity can be expressed as
\begin{equation}
C = \frac{1}{2} \mathbb{E} \left\{ \log_2 \det \left( \mathbf{I}_{N_s} + \frac{P_s}{N_s} \mathbf{U}^\dag \mathbf{A} \mathbf{U} \right)\right\},
\end{equation} 
where $\mathbf{A} = \begin{bmatrix}
F_{{ds}_1} \mathbf{I}_{N_d} & \mathbf{0}\\
\mathbf{0} & \mathbf{\Omega}
\end{bmatrix}$ and $\mathbf{U} = 
\begin{bmatrix}
\mathbf{U}_{dr}^\dag \mathbf{H}_{ds}\\
\mathbf{V}_{dr}^\dag \mathbf{H}_{rs}
\end{bmatrix}$, with $\mathbf{\Omega} = F_{{ds}_2}^2 \mathbf{D}^\dag {\left( F_{dr}^2 \mathbf{D}\mathbf{D}^\dag + \mathbf{I}\right)}^{-1} \mathbf{D}$. 
Note that $\mathbf{U}_{dr}$ and $\mathbf{V}_{dr}$ are unitray matrices, and so do not change the statistics of $\mathbf{H}_{ds}$ and $\mathbf{H}_{rs}$ .


\section{Main Results}

\subsection{Definition of Coverage Region}

The geometry of MIMO relay channel is depicted in Fig.~\ref{Fig2}. In this configuration the source node is located at $d_S = \left(0,0\right)$, the relay node is located at $d_{R} = \left(r_{R}, 0\right)$, and the destination node is located at $d_D = \left(r_D, \theta_D\right)$. We let $\alpha$ be the path loss component. Thus, the channel gain matrices can be written as
\begin{align}\label{pathloss.H}
    \mathbf{H}_{rs} &= \frac{1}{r_R^{\sfrac{\alpha}{2}}}\mathbf{\hat{H}}_{rs}, \; \; \;  
    \mathbf{H}_{ds} = \frac{1}{r_D^{\sfrac{\alpha}{2}}}\mathbf{\hat{H}}_{ds}, \; \; \;
    \mathbf{H}_{dr} &= \frac{1}{{r_{DR}}^{\sfrac{\alpha}{2}}} \mathbf{\hat{H}}_{dr}, \;
\end{align}
where the entries of $\mathbf{\hat{H}}_{rs}$, $\mathbf{\hat{H}}_{ds}$, and $\mathbf{\hat{H}}_{dr}$ are i.i.d. $\mathcal{CN}(0,1)$, and consequently, $\mathbf{\hat{H}}_{rs} \mathbf{\hat{H}^\dagger}_{rs}$, $\mathbf{\hat{H}}_{ds} \mathbf{\hat{H}^\dagger}_{ds}$, and $\mathbf{\hat{H}}_{dr} \mathbf{\hat{H}^\dagger}_{dr}$ are central complex Wishart matrices with
identity covariance matrix \cite{tulino2004random}. 
Furthermore, we assume that each block's use of the channel corresponds to an independent realization of channel matrices.

Now, we define the concept of coverage region as \cite{aggarwal2009maximizing,razeghi2014iswcs} 
\begin{align}\label{def.coverage}
    \mathcal{C}\! \mathit{o}\mathit{v} \left(r_R\right) = \left\{d_D : C\left(r_R,d_D\right) \geq R_c \right\},
\end{align}
where $R_c > 0$ denotes the desired transmission rate and $C$ denotes the capacity at relay location $r_R$ and destination location $d_D$.


\subsection{Desired Transmission Rate Analysis}

In order to obtain a theoretical expression between the desired transmission rate $R_c$ and the optimal relay location $d^* = r_R^*$, by using the coverage region definition given by (\ref{def.coverage}) and substituting {\em effective} SNR ($\mathrm{SNR_{eff}} \triangleq P_s . {r_R^{*}}^{- \alpha}$) into (\ref{ergodicAFbyH}), $R_c$ can be defined as
\begin{equation}
R_c \triangleq \frac{1}{2} \mathbb{E} \left\{ \log_2 \det \left( \mathbf{I}_{2N_d} + \frac{P_s}{N_s} {r_R^{*}}^{- \alpha} \mathbf{H}\mathbf{H}^\dag {\left(\mathbf{B} \mathbf{B}^\dag\right)}^{-1} \right)\right\}.
\end{equation}
The definition of $R_c$ stems from the fact that when the relay station is situated at distances smaller than $r_R^{*}$ the coverage region decreases with the reduction of the distance between the relay node and source node, whereas for distances larger than  $r_R^{*}$ coverage region shrinks because of failure to achieve the required quality of service (QoS) at the relay station. Consequently, intending to maximize coverage region,   $r_R^{*}$ can be opted as the optimal relay location.

The desired transmission rate expressed above involves the expectation operator. In order to obtain exact expression for the desired transmission rate, we use the exact ergodic capacity of AF MIMO relay channel with direct link as \cite{firag2009afcapacitydirect} 
\begin{align}
R_c = \sum_{i=q-s+1}^{q} \sum_{j=1}^{q} \frac{s}{2 \mathrm{ln}\left(2\right)} C_1\, \left(-1\right)^{i+j} \vert \mathbf{K}_{i,j} \vert . I_B\, ,
\end{align}
where $q = N_d + r$, $s = \min \left(N_s, q\right)$, and $C_ 1$, $\mathbf{K_{i,j}}$ and $I_B$ are functions of channels and transmitters parameters (eigen values, number of antennas, amplification factor, etc.).

\begin{figure}
    \centering
    \includegraphics[width=8.3cm,height=5.1cm]{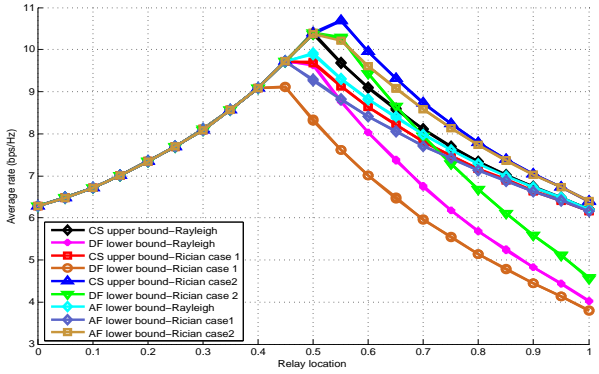}
    \caption{Comparison of capacity bounds with source-relay LOS scenario for $P_s=P_r=10$ dB, $d_y = 0.1$, and $N_a=2$.}
    \label{Fig3}
\end{figure}
\begin{figure}
\centering
\includegraphics[width=8.3cm,height=5.1cm]{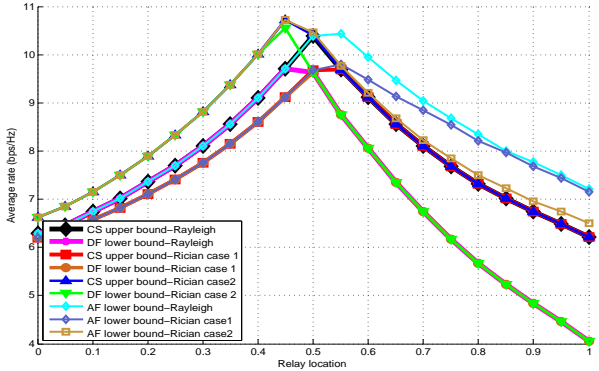}
\caption{Comparison of capacity bounds with relay-destination LOS scenario for $P_s=P_r=10$ dB, $d_y = 0.1$, and $N_a=2$.}
\label{Fig4}
\end{figure}


\section{Simulation Results}

In this section, we confirm our evaluation by using Monte Carlo simulations. By considering three different fading models for source-relay and relay-destination link, we study the effect of line-of-sight and non line-of-sight propagation environment on the capacity bounds and coverage region of AF MIMO relay channel. In order to compare two popular strategies, amplify-and-forward and decode-and-forward, we assume the same conditions as in \cite{alizadeh2012analysis} which were applied for obtaining the results of DF strategy. In our simulations, we assume that $N_s=N_r=N_d=N_a$, $P_s=P_r=10 \, \mathrm{dB}$, $g=\sqrt{\frac{1}{\left(\
N_r+P_{sr}N_r\right)}}$, and path-loss component is $\alpha = 3.56$.

To study the influence of the LOS propagation environment on capacity bounds and coverage region, we consider two situations for relay station: (1) the situation in which the relay is placed above the rooftop, and so the source-relay link can be modeled as the LOS propagation environment channel; (2) the situation in which the relay is placed below the rooftop (on street placement), and so in this case the relay-destination link is modeled by the LOS propagation channel. 

\begin{figure}
    \centering
    \includegraphics[width=8.2cm,height=5.0615cm]{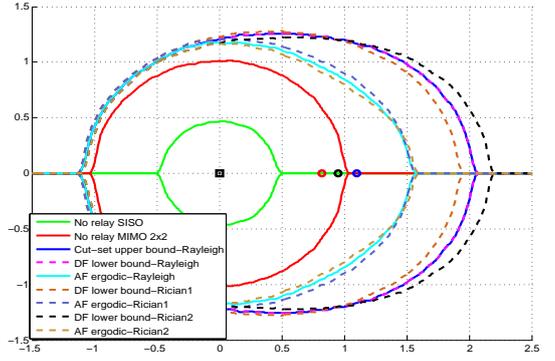}
    \caption{Comparison of coverage region with source-relay LOS scenario when the relay is placed at $r_R^{\ast} = 0.82, 0.95, 1.1$ and $N_a = 2$.}
    \label{Fig5}
\end{figure}
\vspace{0.5pt}
\begin{figure}
\centering
\includegraphics[width=8.2cm,height=5.0615cm]{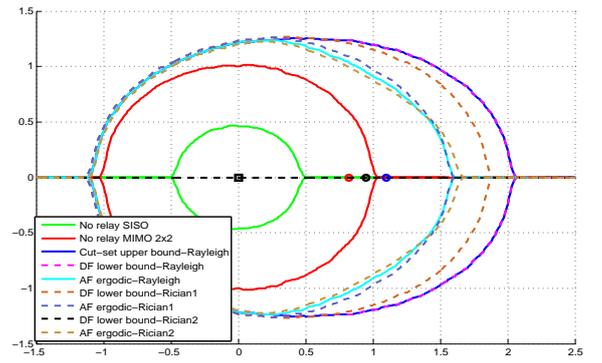}
\caption{Comparison of coverage region with relay-destination LOS scenario when the relay is placed at $r_R^{\ast} = 0.82, 0.95, 1.1$ and $N_a = 2$. }
\label{Fig6}
\end{figure}

Ignoring phase factors, we consider two different LOS components for our channel model as follows\cite{paulraj2003introduction}
\begin{align}
\label{losmatrices}
\mathbf{H}_{1}^{\textrm{LOS}}=
\begin{bmatrix}
1& 1\\
1& 1
\end{bmatrix}, \qquad
\mathbf{H}_{2}^{\textrm{LOS}}=
\begin{bmatrix}
1& -1\\
1& 1
\end{bmatrix},
\end{align}
where $\mathbf{H}_{1}^{\mathrm{LOS}}$ corresponds to unwell-conditioned channel (Rician fading case 1) and $\mathbf{H}_{2}^{\mathrm{LOS}}$ corresponds to well-conditioned channel (Rician fading case 2). The $\mathbf{H}_{1}^{\mathrm{LOS}}$ occurs when transmitter-receiver distance is much greater than the element separation at transmitter antennas, whereas the $\mathbf{H}_{2}^{\mathrm{LOS}}$ occurs when transmitter-receiver distance is comparable to the element separation at transmitter or receiver antennas.

Fig.~\ref{Fig3} and Fig.~\ref{Fig4} depict the effect of channel fading on the capacity bounds for amplify-and-forward and decode-and-forward strategies, with two relay station situation. As it can be seen in these figures, we assume that the source and destination are located at $(0,0)$ and $(1,0)$, respectively, and the relay is located at $(d_x,d_y)$; all in Cartesian coordinates, where $d_y=0.1$ and $d_x$ is changing from 0 to 1.

These figures confirm our argument for choosing optimal relay location, and also shows the significant effect of the channel fading on the capacity bounds. As it is shown in these figures, since the Rician fading Case 2 channel is full rank, and the Rician fading case 1 is rank-deficient, the $\mathbf{H}^{\mathrm{LOS}}_2$ channel outperforms the $\mathbf{H}^{\textrm{LOS}}_1$ channel.

In Fig.~\ref{Fig5} and Fig.~\ref{Fig6}, we compare coverage region for amplify-and-forward and decode-and-forward schemes with three different fading models, and also with two situations for the relay station.  In order to compare our results with those obtained in \cite{alizadeh2012analysis}, we simulate the configuration with the same parameters used in \cite{alizadeh2012analysis}. The respective optimal relay locations $r^*_R$ for Rayleigh fading, Rician fading case 1, and Rician fading case 2 for decode-and-forward scheme are 1, 0.87, and 1.15. We placed the relays at slightly shorter distances, i.e., 0.95, 0.82, and 1.1. 
 
Fig.~\ref{Fig5} shows our results for the situation in which the relay is placed above the rooftop, while Fig.~\ref{Fig6} corresponds to the situation in which the relay station is placed below the rooftop. As it can be seen from the figures, the coverage region of AF MIMO relay channel is smaller than the coverage region of DF MIMO relay channel. The reasoning behind this observation is that rate of AF scheme is low compared to DF scheme. In addition, AF scheme amplifys the noise as well as the signal, which results in reduction of the coverage region. It is obvious that the Rician fading case 2 provides  larger coverage  region than  the Rician fading case 1. Moreover, from these figures it can be interpreted that LOS propagation environment makes a significant impact on the coverage region of DF scheme, while the impact of LOS environment on the coverage region of AF scheme is negligible. 

\begin{figure}
\centering
\includegraphics[width=8.4cm,height=5.1cm]{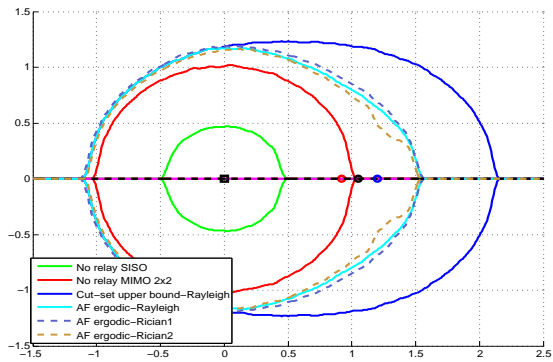}
\caption{Comparison of coverage region with source-relay LOS scenario when the relay is placed at $r_R^{\ast} = 0.92, 1.05, 1.2$ and $N_a = 2$.}
\label{Fig7}
\end{figure}

Fig.~\ref{Fig7} compares the coverage region for relay distances bigger than optimal relay location for DF scheme, i.e., 1.05, 0.92, and 1.2. As the results suggest, in decode-and-forward scheme after a specific distance the coverage region reduces sharply, while in AF scheme coverage region decreases smoothly.

\section{Conclusion}

In this paper, we analyzed coverage region for amplify-and-forward MIMO relay channel with mixed Rician and Rayleigh propagation environment. Considering three fading models with two relay station locations, capacity bounds, coverage region and optimal relay location were obtained. Also, we compared our results with those previous results obtained for decode-and-forward MIMO relay channel with the same parameters. 



%
%



\bibliographystyle{IEEEtran}
\bibliography{references}




\end{document}